\documentclass[twocolumn,prl,showpacs]{revtex4}
\usepackage{amsmath}
\usepackage{graphicx}

\newcommand{\eq}{\begin{equation}}
\newcommand{\fine}{\end{equation}}

\begin{document}

\title{Schroedinger Cat:\\Entanglement test in a Micro-Macroscopic system}
\author{Francesco De Martini$^{1,2}$, Fabio Sciarrino $^{3,1}$, and Chiara Vitelli$%
^{1} $}

\address{$^{1}$Dipartimento di Fisica dell'Universit\'{a} ''La Sapienza'' and Consorzio Nazionale Interuniversitario per le Scienze Fisiche della Materia, Roma, 00185 Italy\\
$^{2}$ Accademia Nazionale dei Lincei, via della Lungara 10, I-00165 Roma, Italy \\
$^3$Centro di Studi e Ricerche ''Enrico Fermi'', Via Panisperna 89/A,Compendio del Viminale, Roma 00184, Italy}

\begin{abstract}
A Macro-state consisting of $\mathbf{N}\approx 3.5\times 10^{4}$ photons in
a quantum superposition and entangled with a far apart single-photon state
(Micro-state) is generated. Precisely, an entangled photon pair is created
by a nonlinear optical process, then one photon of the pair is injected into
an optical parametric amplifier (OPA)\ operating for any input polarization
state, i.e.\ into a phase-covariant cloning machine. Such transformation
establishes a connection between the single photon and the multi particle
fields. We then demonstrate the non-separability of the bipartite system by
adopting a local filtering technique within a positive operator valued
measurement.
\end{abstract}

\maketitle

In recent years two fundamental aspects of quantum mechanics have
attracted a great deal of interest, namely the investigation on
the irreducible nonlocal properties of Nature implied by quantum
entanglement and the physical realization of the ``Schr\oe dinger
Cat'' \cite{Eins35,Schr35}. The last concept, by applying the
nonlocality property to a combination of a microscopic and of a
Macroscopic systems, enlightens the concept of the quantum state,
the dynamics of large systems and ventures into the most
intriguing philosophical problem, i.e. the emergence of quantum
mechanics in the real life. In recent years quantum entanglement
has been demonstrated within a two photon system \cite{Kwia95},
within a single photon and atomic ensemble \cite{Mats05,deRi06}
and within atomic ensembles \cite {Juls01,Moeh07,Chou05}. 
While, according to the 1935 proposal the nonlocal correlations
were conceived to connect the dynamics of two ``microscopic''
objects, i.e. two spins within the well known EPR-Bohm scheme 
\cite{Kwia95}, in the present work the
entanglement is established between a ``Microscopic'' and a
``Macroscopic'', i.e. multi-particle quantum object, via cloning
amplification: Fig. 1. The amplification is achieved by adopting a
high-gain nonlinear (NL) parametric amplifier acting on a
single-photon input carrier of quantum information, i.e., a qubit
state: $\left| \phi \right\rangle $. This process, referred to as
``quantum injected optical parametric amplification''
(\textrm{QI-OPA})\ \cite{DeMa98,DeMa05b} turned out to be
particularly fruitful in the recent past to gain insight into
several little explored albeit fundamental, modern aspects of
quantum information, as \textsl{optimal} quantum cloning machines
\cite{DeMa05b,Pell03,DeMa05}, \textsl{optimal} quantum U-NOT gate
\cite{DeMa02}, quantum no-signaling \cite{DeAn07}. Here, by
exploiting the amplification process, we convert by a unitary
transformation a single photon qubit into a single Macro-qubit
involving a large number of photons, typically $5\times 10^{4}$.
At variance with the previous works \cite{Naga07,DeAn07}, here we
demonstrate for the first time the entanglement between the
microscopic qubit and the macroscopic one obtained by the
amplification process. This result is achieved performing a local
dichotomic measurement on the multiphoton field. Let us venture in
a more detailed account of our endeavor.

\begin{figure}[t]
\includegraphics[scale=.3]{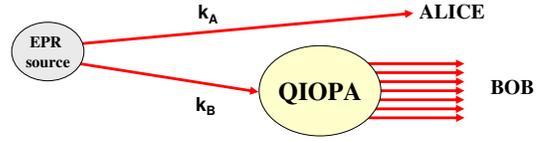}
\caption{Schematic diagram showing the single photon
Quantum-Injected Optical Parametric Amplification (QI-OPA).}
\end{figure}

An entangled pair of two photons in the singlet state $\left| \Psi
^{-}\right\rangle _{A,B}$=$2^{-{\frac{1}{2}}}\left( \left| H\right\rangle
_{A}\left| V\right\rangle _{B}-\left| V\right\rangle _{A}\left|
H\right\rangle _{B}\right) \ $was produced through a Spontaneous Parametric
Down-Conversion (SPDC)\ by the NL\ crystal 1 (C1)\ pumped by a pulsed UV
pump beam: Fig.2. There $\left| H\right\rangle $ and $\left| V\right\rangle $
stands, respectively, for a single photon with horizontal and vertical
polarization while the labels $A,B$ refer to particles associated
respectively with the spatial modes $\mathbf{k}_{A}$ and $\mathbf{k}_{B}$.
Precisely, $A,B$ represent the two space-like separated Hilbert spaces
coupled by the entanglement. The photon belonging to $\mathbf{k}_{B}$,
together with a strong ultra-violet (UV) pump laser beam, was fed into an
optical parametric amplifier consisting of a NL crystal 2 (C2)\ pumped by
the beam $\mathbf{k}_{P}^{\prime }$. The crystal 2, cut for collinear
operation, emitted over the two modes of linear polarization, respectively
horizontal and vertical associated with $\mathbf{k}_{B}$. The interaction
Hamiltonian of the parametric amplification $\widehat{H}=i\chi \hbar
\widehat{a}_{H}^{\dagger }\widehat{a}_{V}^{\dagger }+h.c.$ acts on the
single spatial mode $\mathbf{k}_{B}$ where $\widehat{a}_{\pi }^{\dagger }$
is the one photon creation operator associated with the polarization $%
\overrightarrow{\pi }$. The main feature of this Hamiltonian is its property
of ``phase-covariance'' for ``equatorial'' qubits $\left| \phi \right\rangle
$, i.e. representing equatorial states of polarization, $\overrightarrow{\pi
}_{\phi }=2^{-1/2}\left( \overrightarrow{\pi }_{H}+e^{i\phi }\overrightarrow{%
\pi }_{V}\right) ,\overrightarrow{\pi }_{\phi \perp }=\overrightarrow{\pi }%
_{\phi }^{\perp }$, in a Poincar\'{e} sphere representation having $%
\overrightarrow{\pi }_{H}$ and $\overrightarrow{\pi }_{V}$ as the opposite
''poles'' \cite{Naga07}. The equatorial qubits are expressed in terms of a
single phase $\phi \in (0,2\pi )$ in the basis $\left\{ \left|
H\right\rangle ,\left| V\right\rangle \right\} $. The overall output state
amplified by the OPA apparatus is expressed, in any polarization equatorial
basis $\left\{ \overrightarrow{\pi }_{\phi },\overrightarrow{\pi }_{\phi
\perp }\right\} $, by the Micro-Macro entangled state \cite{Schl01}:

\begin{equation}
\left| \Sigma \right\rangle _{A,B}=2^{-1/2}\left( \left| \Phi ^{\phi
}\right\rangle _{B}\left| 1\phi ^{\perp }\right\rangle _{A}-\left| \Phi
^{\phi \perp }\right\rangle _{B}\left| 1\phi \right\rangle _{A}\right)
\label{outputstate}
\end{equation}
where the mutually orthogonal multi-particle ``Macro-states'' are: {\small
\begin{eqnarray}
\left| \Phi ^{\phi }\right\rangle _{B} &=&\sum\limits_{i,j=0}^{\infty
}\gamma _{ij}\frac{\sqrt{(1+2i)!(2j)!}}{i!j!}\left| (2i+1)\phi ;(2j)\phi
^{\perp }\right\rangle _{B} \\
\left| \Phi ^{\phi \perp }\right\rangle _{B} &=&\sum\limits_{i,j=0}^{\infty
}\gamma _{ij}\frac{\sqrt{(1+2i)!(2j)!}}{i!j!}\left| (2j)\phi ;(2i+1)\phi
^{\perp }\right\rangle _{B}
\end{eqnarray}
} with $\gamma _{ij}\equiv C^{-2}(-\frac{\Gamma }{2})^{i}\frac{\Gamma }{2}%
^{j}$, $C\equiv \cosh g$, $\Gamma \equiv \tanh g$, being $g$\ the NL\ gain
\cite{DeMa02}. There $\left| p\phi ;q\phi ^{\perp }\right\rangle _{B}$
stands for a Fock state with $p$ photons with polarization $\overrightarrow{%
\pi }_{\phi }$ and $q$ photons with $\overrightarrow{\pi }_{\phi \perp }$
over the mode $\mathbf{k}_{B}$. Most important, any injected single-particle
qubit $(\alpha \left| \phi \right\rangle _{B}+\beta \left| \phi ^{\bot
}\right\rangle _{B})$ is transformed by the \textit{information preserving}
QI-OPA\ operation into a corresponding Macro-qubit $(\alpha \left| \Phi
^{\phi }\right\rangle _{B}+\beta \left| \Phi ^{\phi \perp }\right\rangle
_{B})$ macroscopic quantum superposition
\cite{DeMa98}. The quantum states of Eq.(2-3) deserve some comments. The
multi-particle states $\left| \Phi ^{\phi }\right\rangle _{B}$, $\left| \Phi
^{\phi \perp }\right\rangle _{B}$ are orthonormal and exhibit observables
bearing macroscopically distinct average values. Precisely, for the
polarization mode $\overrightarrow{\pi }_{\phi }$ the average number of
photons is $\overline{m}=\sinh ^{2}g$ for $\left| \Phi ^{\phi \perp
}\right\rangle_{B}$, and $(3\overline{m}+1)$ for $\left| \Phi ^{\phi
}\right\rangle _{B}$. For the $\pi -$mode $\overrightarrow{\pi }_{\phi \perp
}$ these values are interchanged among the two Macro-states. On the other
hand, as shown by \cite{DeMa98}, by changing the representation basis from $%
\left\{ \overrightarrow{\pi }_{\phi },\overrightarrow{\pi }_{\phi \perp
}\right\} \;$to $\left\{ \overrightarrow{\pi }_{H},\overrightarrow{\pi }%
_{V}\right\} $, the same Macro-states, $\left| \Phi ^{\phi }\right\rangle
_{B}$ or $\left| \Phi ^{\phi \perp }\right\rangle _{B}$ are found to be
quantum superpositions of two orthogonal states $\left| \Phi
^{H}\right\rangle _{B}$, $\left| \Phi ^{V}\right\rangle _{B}$ which differ
by a single quantum. This unexpected and quite peculiar combination, i.e. a
large difference of a measured observable when the states are expressed in
one basis and a small Hilbert-Schmidt distance of the same states when
expressed in another basis turned out to be a useful and lucky property
since it rendered the coherence patterns of our system very robust toward
coupling with environment, e.g. losses.
The decoherence of our system was investigated experimentally and
theoretically in the laboratory: cfr: \cite{DeMa05,Cami06,Naga07}.

As shown in Figure 2, the single particle field on mode $\mathbf{k}_{A}$ was
analyzed in polarization through a Babinet-Soleil phase-shifter (PS), i.e. a
variable birefringent optical retarder, two waveplates $\left\{ \frac{%
\lambda }{4},\frac{\lambda }{2}\right\} $ and polarizing beam splitter (%
\text{PBS}). It was finally detected by two single-photon detectors $D_{A}$
and $D_{A\text{ }}^{\ast }$ (ALICE box). \ The multiphoton QI-OPA\ amplified
field associated with the mode $\mathbf{k}_{B}$ was sent, through a
single-mode optical fiber (SM),\ to a measurement apparatus consisting of a
set of \ waveplates $\left\{ \frac{\lambda }{4},\frac{\lambda }{2}\right\} $%
, a (PBS) and two photomultipliers (PM) $P_{B}$ and $P_{B}^{\ast }$ (BOB
box). The output signals of the PM's were analyzed by an ``orthogonality
filter'' (OF) that will be described shortly in this paper.\

\begin{figure}[t]
\includegraphics[scale=.30]{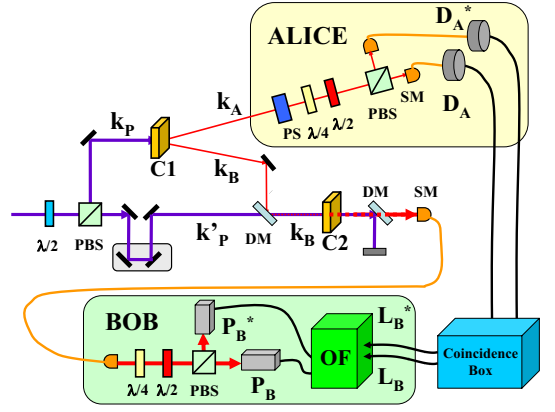}
\caption{Optical configuration of the QI-OPA apparatus. The excitation
source was a Ti:Sa Coherent MIRA mode-locked laser amplified by a Ti:Sa
regenerative REGA device operating with repetition rate $250kHz$. The output
beam, frequency-doubled by second-harmonic generation, provided the OPA\
excitation field beam at the UV wavelength (wl) $\protect\lambda _{P}=397.5nm
$ with power: $750\div 800mW$. A type II BBO crystal (crystal 1: C1)
generates pair of photons with wavelength $\protect\lambda= 2\protect\lambda%
_{p}=795 nm$. C1 generates an average photon number per mode equal to about $%
0.35$, while the overall detection efficiency of the trigger mode was
estimated to be $\simeq 5\%$ with detection rates of about $5kHz$. The NL BBO crystal 2: C2, realizing the
optical parametric amplification (OPA), is cut for collinear type II phase
matching. Both crystals C1 and C2 are 1.5 mm thick. The fields are coupled
to single mode (SM) fibers. The overall detection efficiency on mode $k_{B}$ has been estimated to be $\sim 2\%$.}
\label{outputstate}
\end{figure}

We now investigate the bipartite entanglement between the modes $\mathbf{k}%
_{A}$ and $\mathbf{k}_{B}$. We define the $\frac{1}{2}-$spin Pauli operators
$\left\{ \hat{\sigma}_{i}\right\} $ for a single photon polarization state,
where the label $i$ $=(1,2,3)$ refer to the polarization bases: $%
i=1\Longleftrightarrow \left\{ \overrightarrow{\pi }_{H},\overrightarrow{\pi
}_{V}\right\} $, $i=2\ \Longleftrightarrow \left\{ \overrightarrow{\pi }_{R},%
\overrightarrow{\pi }_{L}\right\} $, $i=3\Longleftrightarrow \left\{
\overrightarrow{\pi }_{+},\overrightarrow{\pi }_{-}\right\} $. Here $%
\overrightarrow{\pi }_{R}=2^{-1/2}(\overrightarrow{\pi }_{H}-i%
\overrightarrow{\pi }_{V}),\overrightarrow{\pi }_{L}=\overrightarrow{\pi }%
_{R}^{\perp }$ \ are the right and left handed circular polarizations and $%
\overrightarrow{\pi }_{\pm }=2^{-1/2}(\overrightarrow{\pi }_{H}\pm
\overrightarrow{\pi }_{V})$. It is found $\hat{\sigma}_{i}=\left| \psi
_{i}\right\rangle \left\langle \psi _{i}\right| -\left| \psi _{i}^{\perp
}\right\rangle \left\langle \psi _{i}^{\perp }\right| $ where $\left\{
\left| \psi _{i}\right\rangle ,\left| \psi _{i}^{\perp }\right\rangle
\right\} $ are the two orthogonal qubits corresponding to the $%
\overrightarrow{\pi }_{i}$ basis, e.g., $\left\{ \left| \psi
_{1}\right\rangle ,\left| \psi _{1}^{\perp }\right\rangle \right\} $ = $%
\left\{ \left| H\right\rangle ,\left| V\right\rangle \right\} $, etc. By the
QI-OPA unitary process the single-photon $\hat{\sigma}_{i}$ operators evolve
into the ``Macro-spin'' operators: $\hat{\Sigma}_{i}=\hat{U}\hat{\sigma}_{i}%
\hat{U}^{\dagger }=\left| \Phi ^{\psi i}\right\rangle \left\langle \Phi
^{\psi i}\right| -\left| \Phi ^{\psi i\perp }\right\rangle \left\langle \Phi
^{\psi i\perp }\right| .$ Since the operators $\left\{ \hat{\Sigma}%
_{i}\right\} $ are built from the unitary evolution of eigenstates of $\hat{%
\sigma}_{i}$ , they satisfy the same commutation rules of the single
particle $\frac{1}{2}-$spin: $\left[ \hat{\Sigma}_{i},\hat{\Sigma}_{j}\right]
=\varepsilon _{ijk}2i\hat{\Sigma}_{k}$ where $\varepsilon _{ijk}$ is the
Levi-Civita tensor density. The generic state $(\alpha \left| \Phi
^{H}\right\rangle _{B}+\beta \left| \Phi ^{V}\right\rangle _{B})$ is a
Macro-qubit in the Hilbert space $B$ spanned by $\left\{ \left| \Phi
^{H}\right\rangle _{B},\left| \Phi ^{V}\right\rangle _{B}\right\} $, as
said. To test whether the overall output state is entangled, one should
measure the correlation between the single photon spin operator $\hat{\sigma}%
_{i}^{A}$ on the mode $\mathbf{k}_{A}$ and the Macro-spin operator $\widehat{%
\Sigma }_{i}^{B}$ on the mode $\mathbf{k}_{B}$. We then adopt the criteria
for two qubit bipartite systems based on the spin-correlation. We define the
``visibility'' $V_{i}=\left| \left\langle \widehat{\Sigma }_{i}^{B}\otimes
\widehat{\sigma }_{i}^{A}\right\rangle \right| $ a parameter which
quantifies the correlation between the systems $A$ and $B.$ Precisely $%
V_{i}=\left| P(\psi _{i},\Phi ^{\psi i})+P(\psi _{i}^{\perp },\Phi ^{\psi
i\perp })-P(\psi _{i},\Phi ^{\psi i\perp })-P(\psi _{i}^{\perp },\Phi ^{\psi
i})\right| $ where $P(\psi _{i},\Phi ^{\psi i})$ is the probability to
detect the systems $A$ and $B$ in the states $\left| \psi _{i}\right\rangle
_{A}$ and $\left| \Phi ^{\psi i}\right\rangle _{B}$, respectively. The value
$V_{i}=1$ corresponds to perfect anti-correlation, while $V_{i}=0$ expresses
the absence of any correlation. The following upper bound criterion for a
separable state holds \cite{Eise04}:
\begin{equation}
S=(V_{1}+V_{2}+V_{3})\leq 1
\end{equation}
In order to measure the expectation value of $\widehat{\Sigma }_{i}^{B}$ a
discrimination among the pair of states $\left\{ \left| \Phi ^{\psi
i}\right\rangle _{B},\left| \Phi ^{\psi i\perp }\right\rangle _{B}\right\} $
for the three different polarization bases $1,2,3$ is required.\ Consider
the Macro-states $\left| \Phi ^{+}\right\rangle _{B}$, $\left| \Phi
^{-}\right\rangle _{B}$ expressed by Equations 2-3, for $\phi =0$\ and $\phi
=\pi $. In principle, a perfect discrimination can be achieved by
identifying whether the number of photons over the $\mathbf{k}_{B}$ mode
with polarization $\overrightarrow{\pi }_{+}$ is even or odd, i.e. by
measuring an appropriate ``parity operator''. This requires the detection of
the macroscopic field by a perfect \ \textit{photon-number resolving}
detectors operating with an overall quantum efficiency $\eta $ $\approx $ 1,
a device out of reach of the present technology.

\begin{figure}[h]
\includegraphics[scale=.26]{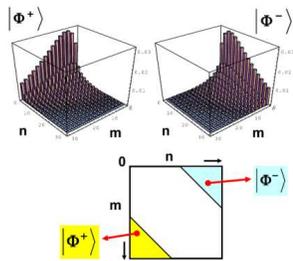}
\caption{Theoretical\ probability distributions $P^{\pm }(m,n)$ of the
number of photons associated with the Macro-states $\left| \Phi ^{\pm
}\right\rangle $ ($g=1.6$). Probabilistic identification of the
wavefunctions $\left| \Phi ^{\pm }\right\rangle $ by OF-filtering the $%
P^{\pm }(m,n)$ distributions over the photon number two-dimensional space $%
\left\{ m,n\right\} $. The white section in the cartesian plane $(m,n)$
corresponds to the ``inconclusive events'' of our POVM OF-filtering
technique.}
\end{figure}

It is nevertheless possible to exploit, by a somewhat sophisticated
electronic device dubbed ``Orthogonality Filter'' (OF), the macroscopic
difference existing between the functional characteristics of the
probability distributions of the photon numbers associated with the quantum
states $\left\{ \left| \Phi ^{\pm }\right\rangle _{B}\right\} $. The
measurement scheme works as follows: Figures 2 and 3. The multiphoton field
is detected by two PM's $(P_{B},P_{B}^{\ast })$ which provide the electronic
signals $(I_{+},I_{-})$ corresponding to the field intensity on the mode $%
\mathbf{k}_{B}$ associated with the $\pi -$components $(\overrightarrow{\pi }%
_{+},\overrightarrow{\pi }_{-})$, respectively. By (OF) the difference
signals $\pm (I_{+}-I_{-})$ are compared with a threshold $\xi k>0$ . When
the condition $(I_{+}-I_{-})>\xi k$ \ is satisfied, the detection of the
state $\left| \Phi ^{+}\right\rangle _{B}$ is inferred and a standard
transistor-transistor-logic (TTL) electronic square-pulse $L_{B}$ is
realized at one of the two output ports of (OF). This corresponds to the
measurement of the eigenvalue $+1$ of the operator $\widehat{\Sigma }_{3}^{B}
$. Likewise, when the condition $(I_{-}-I_{+})>\xi k$ is satisfied, the
detection of the state $\left| \Phi ^{-}\right\rangle _{B}$ is inferred, a
TTL pulse is realized at other output port of (OF) and the eigenvalue of $%
\widehat{\Sigma }_{3}^{B}$ is $-1$. The PM\ output signals are discarded for
$-\xi k<(I_{+}-I_{-})<\xi k$, i.e. in condition of low state discrimination.
By increasing the value of $\ $the\ threshold $k$ an increasingly better
discrimination is obtained together with a decrease of detection efficiency.
This ``local distillation'' procedure is conceptually justified by the
following theorem: since entanglement cannot be created\ or enhanced by any
``local'' manipulation of the quantum state, the non-separability condition
demonstrated for a ``distilled'' quantum system, e.g., after application of
the OF-filtering procedure, fully applies to the same system in absence of
distillation \cite{Eise04}. This statement can be applied to the measurement
of $I_{\phi }$ and $I_{\phi \perp }$ for any pair of quantum states $\left\{
\left| \Phi ^{\phi }\right\rangle _{B},\left| \Phi ^{\phi \perp
}\right\rangle _{B}\right\} $. This method is but an application of a
Positive Operator Value Measurement \ procedure (POVM)\ \cite{Pere95} by
which a large discrimination between the two states $\left\{ \left| \Phi
^{\pm }\right\rangle _{B}\right\} $ is attained at the cost of a reduced
probability of a successful detection.\newline
The present experiment was carried out with a gain value $g=4.4$ leading to
a number of output photons $N\approx 3\times 10^{4}$, after OF filtering. In
this case the probability of photon transmission through the OF\ filter was:
$p\approx 10^{-4}$. A NL gain $g=6$ was also achieved with no substantial
changes of the apparatus. Indeed, an unlimited number of photons could be
generated in principle by the QI-OPA technique, the only limitation being
due to the fracture of the NL crystal 2 in the focal region of the laser
pump. In order to verify the correlations existing between the single photon
generated by the NL\ crystal 1 and the corresponding amplified Macro-state,
we have recorded the coincidences between the single photon detector signal $%
D_{A}$ (or $D_{A}^{\ast }$) and the TTL signal $L_{B}$ (or $L_{B}^{\ast }$)
both detected in the same $\pi -$basis $\left\{ \overrightarrow{\pi }_{+},%
\overrightarrow{\pi }_{-}\right\} $:\ Figure 2. This measurement has\ been
repeated by adopting the common basis $\left\{ \overrightarrow{\pi }_{R},%
\overrightarrow{\pi _{L}}\right\} $. Since the filtering technique can
hardly be applied to the $\left\{ \overrightarrow{\pi }_{H},\overrightarrow{%
\pi }_{V}\right\} $ basis, because of the lack of a broader$\
SU(2)$ covariance of the amplifier, the quantity $V_{1}>0$ could
be measured adopting a photon number resolving detector with
quantum efficiency equal to $1$: a device not made available by the present
technology.
The phase$\ \phi $ between the $\pi -$components $%
\overrightarrow{\pi }_{H}$ and $\overrightarrow{\pi }_{V}$ on mode $\mathbf{k%
}_{A}$ was determined by the Babinet-Soleil variable phase shifter ($PS$)$.$
Figure 4 shows the fringe patterns obtained by recording the rate of
coincidences of the signals detected by the Alice's and Bob's measurement
apparata,\ for different values of $\phi $. These patterns were obtained by
adopting the common analysis basis $\left\{ \overrightarrow{\pi }_{R},%
\overrightarrow{\pi }_{L}\right\} \;$with a filtering probability $\simeq
10^{-4}$, corresponding to a threshold $\xi k$ about $8$ times higher than
the average photomultiplier signals $I$. In this case the average visibility
has been found $V_{2}=(54.0\pm 0.7)\%$. A similar oscillation pattern has
been obtained in the basis $\left\{ \overrightarrow{\pi }_{+},%
\overrightarrow{\pi }_{-}\right\} $ leading to: $V_{3}=\left( 55\pm 1\right)
\%$. Since always is $V_{1}>0$, our experimental result $S\;$= $V_{2}+V_{3}\;
$= $(109.0\pm 1.2)\%$ implies the violation of the separability criteria of
Equation (4) and then demonstrates the non-separability of our Micro-Macro
system belonging to the space-like separated Hilbert spaces $A$ and $B$. By
evaluating\ the experimental value of the ``concurrence'' for our test,
connected with the ``entanglement of formation'', it is obtained $C\geq
0.10\pm 0.02>0$ \cite{Benn96,Hill97}. This result again confirms the
non-separability of our bipartite system.\ The value of $C$ could be increased by improving the matching of the injected photon with the pump field on crystal $C2$.
A method similar to ours to test
the non-separability of a 2-atom bi-partite system was adopted recently by
\cite{Moeh07}.

In conclusion we have demonstrated the entanglement of a Micro-Macro system,
in which the Macro state has been obtained by the amplification of a 1-photon qubit. The QI-OPA approach
could be directly applicable to the field of Quantum Information and
Computation in virtue of the intrinsic \textit{information-preserving} \
property of the QI-OPA\ dynamics. Indeed, this property implies the direct
realization of the quantum map $(\alpha \left| \phi \right\rangle +\beta
\left| \phi ^{\bot }\right\rangle )$ $\longrightarrow $ $(\alpha \left| \Phi
^{\phi }\right\rangle +\beta \left| \Phi ^{\phi \perp }\right\rangle )$
connecting any single-particle qubit to a corresponding Macro-qubit, by then
allowing the extension to the multi particle regime of most binary logic
algorithms and techniques. For instance, let's consider a 2-qubit phase gate in which the
control-target interaction is provided by a Kerr-type optical nonlinearity.
The strength of this nonlinearity is far too small to provide a sizable
interaction between the ``control'' and the ``target'' single-particle
qubits.
However, by replacing these ones by the corresponding Macro-qubits
associated with $N$ photons, the NL interaction strength can be enhanced by
many orders of magnitude since the 3d-order NL\ polarization scales as $%
N^{3/2} $. This application is made possible by another general property
of the QI-OPA scheme, i.e., the direct accessibility of the Macro-states at the output of the QI-OPA.
In summary, the amplification process applied to a Micro system is
a natural approach to enlighten the quantum-to-classical transition and to
investigate the persistence of quantum phenomena into the ``classical''
domain by measurement procedures applied to systems of increasing
size.

\begin{figure}[t]
\includegraphics[scale=.2]{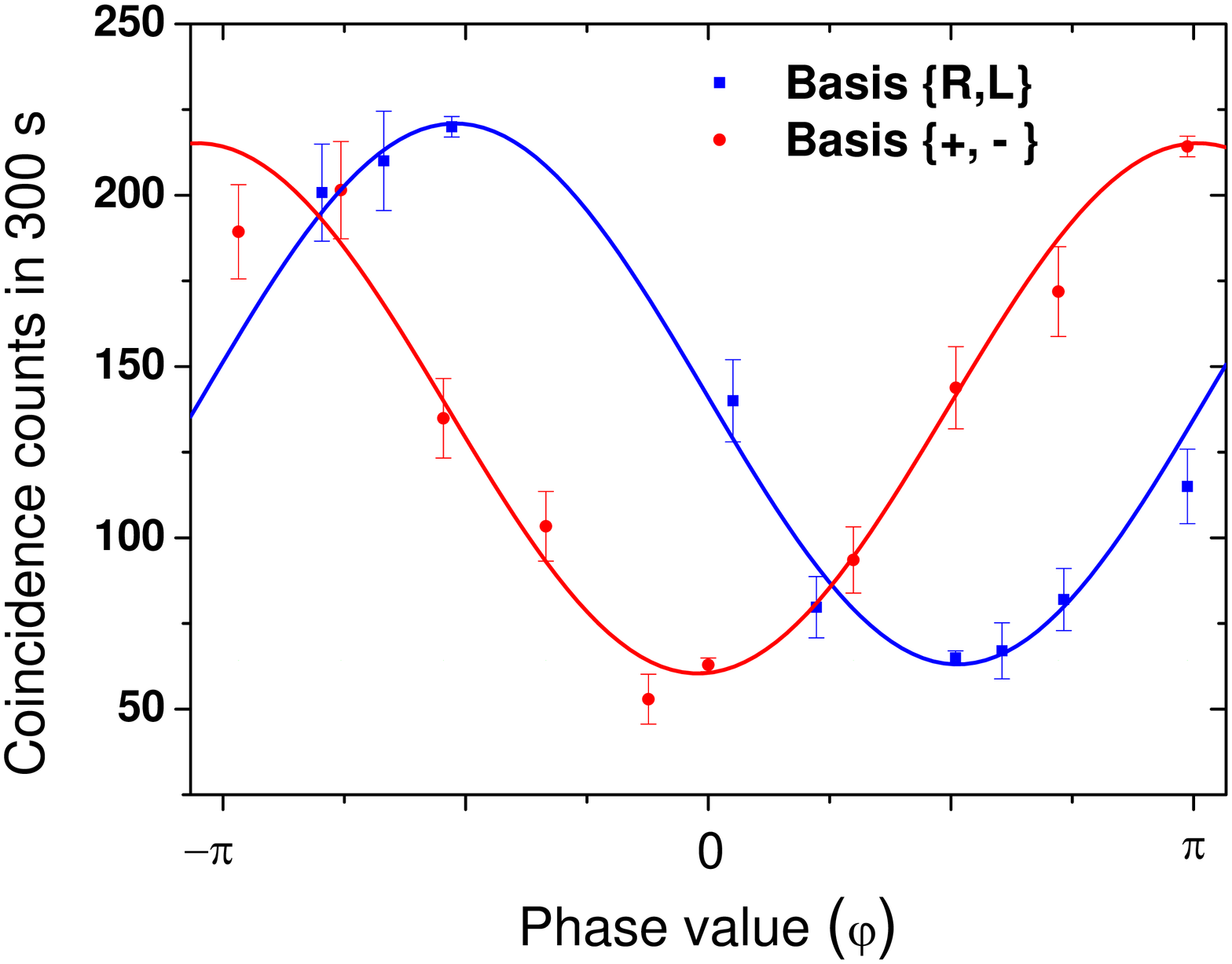}
\caption{Coincidence counts $[L_{B},D_{A}]$ versus the phase $\protect\phi $ of the injected
qubit for the  basis $\left\{ \vec{\protect\pi }_{+},\vec{\protect\pi }%
_{-}\right\}$ (circle data) and the basis $\left\{ \vec{\protect\pi }_{R},\vec{\protect\pi }%
_{L}\right\}$ (square data). The experimental points corresponding to the minima and maxima have been adopted to 
estimate $V_{2}$ and $V_{3}$. Accordingly, they have been determined by a higher statistics and exhibit a smaller
 error flag. }
\end{figure}

We acknowledge support and collaboration from Sandro Giacomini, Giorgio Milani, Eleonora Nagali, Tiziano De Angelis, Nicolo' Spagnolo. Work supported by the PRIN 2005 of MIUR and
INNESCO 2006 of CNISM.

\end{document}